\documentclass{siamart220329}
\usepackage[utf8]{inputenc}

\title{Quantum Navier-Stokes equations \\ for electrons in graphene}
\author{Luigi Barletti\thanks{Dipartimento di Matematica e Informatica ``U.\ Dini'', Universit\`a degli Studi di Firenze, Italy  (\email{luigi.barletti@unifi.it}).} 
\and
Lucio Demeio\thanks{Dipartimento di ingegneria industriale e scienze matematiche, Universit\`a di Politecnica delle Marche, Italy (\email{demeio@dipmat.univpm.it}, \email{s.nicoletti@staff.univpm.it}).}
\and
Sara Nicoletti\footnotemark[3]
}
\usepackage[american]{babel}
\usepackage{bm}
\usepackage{amsfonts}
\usepackage{graphicx}
\usepackage{amsmath}
\usepackage{amssymb}
\usepackage{color}
\usepackage{comment}
\newtheorem{rem}{Remark}
\newcommand{\BE}{\begin{equation}}
\newcommand{\EE}{\end{equation}}
\newcommand{\BEA}{\begin{eqnarray}}
\newcommand{\EEA}{\end{eqnarray}}
\newcommand{\BEN}{\begin{eqnarray*}}
\newcommand{\EEN}{\end{eqnarray*}}
\newcommand{\cE}{\mathcal{E}}

\newcommand{\cO}{\mathcal{O}}

\newcommand{\mR}{\mathbb{R}}
\newcommand{\e}{\mathrm{e}}

\newcommand{\vphi}{\varphi}
\newcommand{\BB}{\boldsymbol{B}} 
\newcommand{\FF}{\boldsymbol{F}} 
\newcommand{\GG}{\boldsymbol{G}} 
\newcommand{\xx}{{\boldsymbol{x}}} 
\newcommand{\yy}{{\boldsymbol{y}}} 
\newcommand{\pp}{{\boldsymbol{p}}}

\newcommand{\kk}{{\boldsymbol{k}}}
\newcommand{\vv}{{\boldsymbol{v}}}
 
\newcommand{\uu}{\boldsymbol{u}} 
\newcommand{\jj}{{\boldsymbol{j}}}
\newcommand{\nnu}{\boldsymbol{\nu}} 
\newcommand{\eeta}{\boldsymbol{\eta}} 
\newcommand{\xxi}{\boldsymbol{\xi}} 
\newcommand{\vecA}{\boldsymbol{A}} 
\newcommand{\vecN}{\boldsymbol{N}}

\newcommand{\bk}[1]{{\left\langle #1 \right\rangle}} 
\newcommand{\abs}[1]{{\vert {#1} \vert}}

\newcommand{\pt}{\partial}

\DeclareMathOperator{\Op}{Op}

\begin{document}
\maketitle
\begin{abstract}
The Chapman-Enskog method, in combination with the quantum maximum entropy principle, is applied to the Wigner equation in order to obtain quantum Navier-Stokes equations for electrons in graphene in the isothermal case.  The derivation is based on the quantum version of the maximum entropy principle and follows the lines of Ringhofer-Degond-M\'ehats' theory ({\it J. Stat. Phys.} \textbf{112}, 2003 and {\it Z. Angew. Math. Mech.} {\bf 90}, 2010). The model  obtained in this way is then semiclassically expanded up to $\cO(\hbar^2)$.
\end{abstract}
%
% REQUIRED

% REQUIRED
\section{Introduction}
\label{S1}
Graphene is a two-dimensional material that displays unusual electronic properties due to the conical shape of the energy bands in the proximity of the so-called Dirac points. For this reason, charge transport in graphene attracts a lot of interest by the scientific community. In the literature, there are several papers describing electron hydrodynamics with semiclassical or quantum models \cite{Barletti14,LucaRomano19,LF18,MascaliRomano17,Muller09,Narozhny19,Zamponi11}.
In particular, the viscous hydrodynamic regime, which is only possible in a time-scale where electron-electron collisions are dominant, has been deeply investigate both theoretically \cite{LF18,Muller09,Narozhny19} and experimentally \cite{Nature17}.
In Refs.\ \cite{LF18,MascaliRomano17,Muller09,Narozhny19}, the author reports a derivation of the viscous hydrodynamic equations based on a semiclassical Boltzmann equation, i.e.\ a kinetc description that is classical except for some key elements such as the conical energy bands and the use of Fermi statistics.
On the other hand, quantum-corrected hydrodynamic equations (obtained from a quantum kinetic description based on the Wigner equation) have been obtained only in the case of inviscid regime and assuming regularized energy bands \cite{LucaRomano19}.
This is because the singular behaviour of the bands at the conical intersection
makes divergent integrals appear when computing quantum corrections to the transport coefficients.
\par
The aim of the present work is to explore the possibility 
of deriving fully-quantum and quantum-corrected hydrodynamic equations in both
the inviscid and viscous regimes.
For the sake of simplicity we shall only consider the isothermal case, since the non-isothermal case would only bring more calculation but no methodological differences.
We shall work with regularized bands, by introducing a regularization parameter $\alpha$ (with the physical meaning of a energy gap), but we shall also consider the conical limit $\alpha \to 0$ whenever possible. 
\par
Quantum hydrodynamic models date back to the early period of quantum mechanics.
The Schr\"odinger equation can be formulated in a hydrodynamic form, the so-called Madelung equations \cite{Madelung}, which represent an isothermal Euler system with an additional $\hbar^2$ term, called Bohm potential \cite{Bohm1,Bohm2}.
Madelung's approach is restricted to pure states, and so it is not suitable to
introduce statistical concepts, such as equilibrium states, which are needed in the derivation of more general quantum fluid models.
A more systematic  way to derive quantum hydrodynamic models was proposed by Degond, Ringhofer and M\'ehats \cite{DMR05,DR03}
by introducing the quantum entropy minimisation principle (QMEP) to close the systems of the macroscopic moments. 
In \cite{DMR05,DR03}, the authors, similarly to the classical maximum entropy principle \cite{CMR2020,Janes80} and to Levermore's method \cite{Levermore}, close the moment system by means of a quantum distribution function which minimizes the entropy with given moments.
This approach is fairly general and can be applied to different systems and different regimes.
\par
Our derivation starts from the one-particle Wigner equation \cite{Wigner,Zachos05} with a BGK collisional term, which describes the relaxation of the system to the local equilibrium.  Here, we are interested in  the isothermal fluid equations for the particle density $n$ and for
the momentum density $n_\pp$. 
Following the the lines of Ringhofer-Degond-M\'ehats' theory,  we obtain the local equilibrium Wigner function by using the QMEP, which determines the local equilibrium by maximizing an entropy functional with the constraints given by the macroscopic moments. 
The equilibrium Wigner function $g$ can be formally derived and it contains an implicit dependence upon the moments $n$ and $n_\pp$ through the dependence on a set of three Lagrange multipliers, $A$ and $\mathbf{B}=(B_1,B_2)$.
The quantum Navier-Stokes equations for graphene are then obtained by performing the limit of the Wigner equation as the mean free path converges to zero, by using the Chapman-Enskog expansion. 
 
At order $0$ in the collision-time parameter we obtain the quantum analogue of the classical Euler equations, while at order $1$ the viscous corrections to the Euler model are obtained. Finally, we determine the expression of macroscopic moments as functions of the Lagrange multipliers in the semiclassical case (namely, at second order in $\hbar$) and then compute the approximated quantum Navier-Stokes equations. 
We also discuss the case of non-regularized (conical) energy
bands, letting the regularization parameter $\alpha$ go to 0, and show that, remarkably, the Euler equations remain non-singular in this limit.
\par
We remark that our approach is analogous to the one used in \cite{BM10}, also based on the QMEP, where the authors derive for the first time quantum isothermal Navier-Stokes equations for a standard particle (i.e.\ with quadratic energy band).
\par
The outline of the paper is the following. 
In Section \ref{S2} we introduce the model recalling the form of the  energy bands for graphene and of their regularization and write down the Wigner equation with a BGK collisional operator. 
In Section \ref{SecCE} we perform the Chapman-Enskog expansion and derive quantum Euler equations and quantum Navier-Stokes equations. In Section \ref{SecSCL} we compute the semiclassical expansion up to $\cO(\hbar^2)$ for the equilibrium distribution and consequently for the Navier-Stokes equations. Section 5 contains our conclusions.
\section{Graphene energy bands and kinetic equations}
\label{S2}
Graphene is a monolayer crystal consisting of a honeycomb lattice of carbon atoms. In the tight binding approximation, the valence and conduction bands 
have conical intersections at the vertices of the Brillouin zone (the so called Dirac points), where they touch each other with 
a null gap  \cite{CastroNeto}.
This leads to consider the carriers as ``massless fermions''. 
In this work, we restrict our model to the conduction band, for which the explicit expression 
\BE
\cE(\kk) = \hbar \, c \, k, \qquad k = \abs{\kk}, \qquad  \label{Ek}
\EE
holds for $\kk$ in a neighborhood of a given Dirac point.
Here, $c$ is the Fermi speed and $\hbar$ is the reduced Planck's constant.
It is customary in the models involving the graphene bands to place the origin of the Brillouin zone at the Dirac point, so the crystal vector $\kk$ in equation \eqref{Ek} is measured from the Dirac point. 
The band profile given by equation \eqref{Ek} presents a singular point at the origin, which causes certain integrals appearing in the kinetic formulations to diverge. However, one can introduce a very small gap related to the first and second neighbour hopping energy; this assumption makes  it possible to use a smooth version of the energy band \cite{CastroNeto,LucaRomano19} given by
\BE
\cE(\kk) = c \sqrt{\hbar^2 \, k^2+\alpha^2},  \label{Eksmooth}
\EE
where $\alpha$ is a small parameter, physically related to the nearest-neighbour and next nearest-neighbour hopping energies. 
\par

Let $w(\xx,\kk,t)$ be the single electron Wigner function  
\cite{Barletti03,JungelBook,Tatarski83,Zachos05}, which is the Wigner
transform of the electron density matrix $\rho(\xx,\yy,t)$
\BE
\label{WiT}
  w(\xx,\kk,t) = \frac{1}{2\pi} \int_{\mR^2} \rho\Big(
  \xx + \frac{\xxi}{2},\xx- \frac{\xxi}{2} \Big)\,\e^{-i\xxi\cdot\kk} d\xxi .
\EE
We suppose that the carrier population lives mainly near the Dirac point of the Brillouin zone, which we have chosen as our origin of the $\kk$ space. This allows to effectively enlarge the Brillouin zone to the whole of $\mR^2$ when integrating over $\kk$.
This entitles us to identify the crystal momentum $\hbar\kk$ with the Wigner canonical momentum $\pp$ so that the Wigner function is viewed as a function of $\pp$ 
(instead of $\kk$),  i.e., $w=w(\xx,\pp,t)$. 
Accordingly, the energy band will be regarded as a function of $\pp$:
\BE
\label{Band}
 \cE(\pp) = c \sqrt{p^2+\alpha^2},
\EE
where $p = \abs{\pp}$. 
A crucial quantity associated with the energy band is the semiclassical velocity, which is defined according to the classical Hamiltonian picture:
\BE
\label{velocity}
\vv(\pp) = \nabla_{\pp} \, \cE(\pp)  = \frac{c}{\sqrt{p^2+\alpha^2}}\,\pp.
\EE

For the sake of simplicity, in this paper we consider an isothermal electron gas, 
so we only need the first two hydrodynamic moments, namely the charge density $n(\xx,t)$ 
and the momentum density $n_\pp(\xx,t)$:
\BEA
&&n(\xx,t) = \int_{\mR^2} w(\xx,\pp,t) \, d\pp \equiv \bk{w} 
\label{density}, 
\\
&&n_\pp(\xx,t) = \int_{\mR^2} w(\xx,\pp,t) \, \pp \, d\pp \equiv \bk{\pp \, w}.
\label{momentum}
\EEA
With a more compact notation, we define the monomials
\BE
 \psi_0(\pp) = 1,\qquad \psi_1(\pp) = p_1, \qquad \psi_2(\pp) = p_2
\EE
and the corresponding moments
\BE
\label{moments}
  n_s = \bk{\psi_s w}, \qquad s = 0,1,2,
\EE
so that 
\BE
n = n_0 \qquad
n_\pp = (n_1,n_2).
\EE

\par
In the Wigner (phase-space) formalism, the operator product translates into the Moyal product between two phase space functions
$u(\xx,\pp)$, $v(\xx,\pp)$:
\BE
\label{MP1}
  u\#v = \sum_{k=0}^\infty \hbar^k u \#_k v ,
\EE
where
\BE
\label{MP2}
  u \#_k v = \frac{1}{(2i)^k} \sum_{\abs{\alpha}+\abs{\beta} = k}\frac{(-1)^\abs{\alpha}}{\alpha! \beta!}
  \nabla_{\xx}^\alpha\nabla_{\pp}^\beta u \, \nabla_{\pp}^\alpha\nabla_{\xx}^\beta v.
\EE
It is important to notice that 
\BE
\label{MP3}
  u \#_k v = (-1)^k  v \#_k u.
\EE 
 The time evolution of a single particle is governed by the Wigner equation \cite{LucaRomano19}
\BE
\label{WE}
\pt_t w + \Lambda[\cE] w + \Theta[V] w = \frac{1}{\tau}(g - w).
\EE

The left-hand side of this equation accounts for the Hamiltonian dynamics 
of the electron and is equivalent to the Schr\"odinger equation for mixed states (also called von Neumann equation).
Here, $\Lambda[\cE]$ and $\Theta[V]$ are pseudo-differential operators describing the effect of
the crystal lattice and the external potential $V(\xx)$.
In terms of the Moyal product we have \cite{JSP12}:
\BE
\begin{aligned}
  \Lambda[\cE] w = \frac{i}{\hbar}\{ \cE, w\}_\# &= \frac{i}{\hbar}(\cE\# w - w \# \cE)  \\
  &= \sum_{k=0}^\infty \frac{(-1)^k\hbar^{2k}}{2^{2k}}\sum_{\abs{\alpha} = 2k+1}\frac{1}{\alpha!}
  \nabla_{\pp}^\alpha \cE \, \nabla_{\xx}^\alpha w \, ,
  \label{pseudoE}
\end{aligned}
\EE
\BE
\begin{aligned}
  \Theta[V] w = \frac{i}{\hbar}\{V, w\}_\# &= \frac{i}{\hbar}(V\# w - w \# V)
  \\
  &=  -\sum_{k=0}^\infty  \frac{(-1)^k\hbar^{2k}}{2^{2k}} \sum_{\abs{\alpha} = 2k+1}\frac{1}{\alpha!}
  \nabla_{\xx}^\alpha V\, \nabla_{\pp}^\alpha w
  \label{pseudoV}
\end{aligned}
\EE
(note that our sign convention for $\Theta$ is opposite to that of Ref.\ \cite{LucaRomano19}).
In the following we will also encounter the symmetric version of $\Theta[V]$, namely
\BE
  \Theta_+[V] w = \frac{1}{2}(V\# w + w \# V)
=  \sum_{k=0}^\infty  \frac{(-1)^k\hbar^{2k}}{2^{2k}} \sum_{\abs{\alpha} = 2k}\frac{1}{\alpha!}
  \nabla_{\xx}^\alpha V\, \nabla_{\pp}^\alpha w \, .
  \label{Thetaplus}
\EE
Definitions (\ref{pseudoE}) and (\ref{pseudoV}) in terms the integral representation become:
\BEA
&& \Lambda[\cE] \, w = \frac{i}{\hbar} \, \frac{1}{(2 \, \pi)^2} \int \delta\cE (\pp,\nnu) \, \widetilde{w} (\nnu,\pp) \, e^{i \, \nnu \cdot \xx} d \nnu \, ,
\label{pseudoEF}
 \\
&& \Theta[V] \, w = \frac{i}{\hbar} \, \frac{1}{(2 \, \pi)^2} \int \delta V (\xx,\eeta) \, \widehat{w} (\xx,\eeta) \, e^{i \, \eeta \cdot \pp} d \eeta \, ,
\label{pseudoVF}
\EEA
where $\widetilde{w} (\nnu,\pp)$ is the Fourier transform of $w(\xx,\pp)$ with respect to $x$ and $\widehat{w} (\xx,\eeta)$ is the Fourier transform of 
$w(\xx,\pp)$ with respect to to $\pp$, with the symbols
$$
\begin{aligned}
&\delta\cE (\pp,\nnu) = 
\cE \Big(\pp+\frac{\hbar \, \nnu}{2}\Big) - \cE \Big(\pp-\frac{\hbar \, \nnu}{2}\Big) , 
\\[4pt]
&\delta V (\xx,\eeta) = 
V \Big(\xx+\frac{\hbar \, \eeta}{2}\Big) - V \Big(\xx-\frac{\hbar \, \eeta}{2}\Big) ,
\end{aligned}
$$
for $\xx$, $\pp$, $\nnu$ $\in \mathbb{R}^2$. 
\par
\smallskip

The right-hand side of Eq.\ \eqref{WE} accounts for the effect of collisions.
According to Ringhofer-Degond-M\'ehats' theory \cite{BM10,DMR05,DR03}, 
collisions in a quantum system can be introduced by means of an operator of BGK type, 
describing the relaxation of the system towards a local equilibrium $g$ 
in a characteristic time $\tau$. 
The classical BGK operators share some aspects with the Boltzmann collision operator, 
like for example local conservation of mass, momentum and energy.
What changes in the quantum case is that the local equilibrium state $g$ is represented by 
a suitable Wigner function that maximizes quantum entropy.
Since we are considering isothermal hydrodynamics, only mass and momentum conservation
will be imposed.
The local equilibrium Wigner function $g$ will be described in detail in Section \ref{S3.2}.

\section{The Chapman-Enskog expansion}
\label{SecCE}
In this section, we derive formally the hydrodynamic equations in a fully quantum picture by using the Chapman-Enskog procedure (see \cite{cercignani} for the classical case and, for example, \cite{DR03} for the quantum case). 
Chapman-Enskog theory is an important step in the  formulation of a macroscopic description starting from a microscopic theory.
\subsection{Chapman-Enskog expansion and moments}
\label{S3.1}

The starting point of the Chapman-Enskog method is the formal expansion of the Wigner function
in a series of powers of the collision time $\tau$:
\BE
\label{CE}
 w = w_0 + \tau w_1 + \tau^2 w_2 + \cdots.
\EE
As we said before, we suppose to work in the isothermal case and that the moments $n_s$ 
(i.e.\ $n$ and $n_\pp$) are locally conserved hydrodynamic quantities, so we assume that
\BE
  n_s = \bk{\psi_s w} = \bk{\psi_s g}, \qquad s = 0,1,2. 
\label{CEmom}
\EE
(a condition that guarantees that the BGK operator conserves $n$ and $n_\pp$).

The second step of the Chapman-Enskog procedure is to assume that the time-derivative of the macroscopic moments
(not the macroscopic momemts themselves) can be expanded in powers of $\tau$
\BE
\label{eqn:sviluppot}
\partial_t n_s = \partial^0_t n_s +\tau\partial^1_t n_s + \cdots.
\EE
By substituting the expansion \eqref{CE} in the Wigner equation \eqref{WE}, we obtain
\BEA
&& w_0 = g
\label{wExpansion1}
\\
&& w_1 = - Tw_0 - \pt_t^0 w_0 = - Tg - \pt_t^0 g \label{wExpansion2}
\EEA 
where $  T =  \Lambda[\cE] + \Theta[V]$ and $\pt_t^0$ stands for the time-derivatives of $g$ approximated at order 0, as it results from the expansion \eqref{eqn:sviluppot}.
Indeed, the function $g$ depends on time only through the dependence on the macroscopic quantities $n$ and $n_\pp$, so that 
\BE
\label{eqn:partialt}
\partial_t g = \sum_{s=0}^2 \frac{\delta g}{\delta n_s} \, \frac{\partial n_s}{\partial t} 
= \sum_{s=0}^2\frac{\delta g}{\delta n_s} \left(  \partial^0_t n_s +\tau\partial^1_t n_s + \cdots. \right).
\EE
Note that we used the symbol of functional derivative because, as we shall see, the dependence of $g$ 
on the moments $n_s$ is deeply non-local.
Also note that equations \eqref{CEmom} and \eqref{wExpansion1} entail 
$$
\bk{\psi_s w_k} = 0, \quad s = 0,1,2, \quad k \geq 1.
$$
Multiplying the Wigner equation \eqref{WE} by $\psi_s$ 
and integrating over $\pp$ we obtain
\BE
  \pt_t n_s + \bk{\psi_s T w_0} + \tau \bk{\psi_s Tw_1} = \cO(\tau^2).
\EE
Then, by using expression (\ref{wExpansion2}), we obtain up to $\cO(\tau^2)$,
\BE
\label{NS1}
  \pt_t n_s + \bk{\psi_s T g} = \tau \bk{\psi_s TTg} + \tau \bk{\psi_s T \pt_t^0 g}, \qquad s = 0,1,2 .
\EE
We have therefore identified 
\BE
\partial_t^0 n_s = - \bk{\psi_s T g},
\qquad
\partial_t^1 n_s = \bk{\psi_s TTg} + \bk{\psi_s T \pt_t^0 g}.
\EE
The $\cO(1)$ equations,
\BE
\label{EE1}
  \pt_t n_s + \bk{\psi_s T g} = 0, \qquad s = 0,1,2,
\EE
are the equations of the inviscid hydrodynamics and are the quantum analogue of the Euler equations of classical hydrodynamics, while the $\cO(\tau)$ equations \eqref{NS1} give the quantum analogue of the classical Navier-Stokes equations. 
The meaning of $\pt_t^0$ is that the time-derivatives of moments are given  by Euler equations \eqref{EE1}. 
\subsection{Equilibrium}
\label{S3.2}
In order to proceed with the Chapman-Enskog expansion, the equilibrium distribution $g$ must be made explicit.
As already mentioned in Sec.\ \ref{S2}, $g$ is a Wigner function that maximizes
the QMEP. 
More precisely, according to the QMEP, $g$ is 
the (unique) minimizer of the von Neumann entropy subject to the
constraint that the observed macroscopic moments $n$ and $n_\pp$ are determined by
Eq.\ \eqref{moments}.
We do not repeat here the derivation of the form of $g$ (which can be found e.g.\ 
in Refs.\ \cite{JSP12,DMR05,DR03}) and we limit ourselves to state the result
that one obtains assuming that the entropy is compatible with Fermi-Dirac statistics:
\BE
\label{gdef}
    g = \Op^{-1} \left( \frac{1}{\e^{\Op(\cE - \BB\cdot\pp - A)}+1} \right).
\EE
Here, $\Op$ represents the Weyl quantization and is defined, for a phase space function $a=a(\xx,\pp)$, by
\BE
\label{WeT}
[\Op(a) \vphi](\xx) = \frac{1}{(2 \, \pi \, \hbar)^2} \, \int a\left(\frac{\xx+\yy}{2},\pp\right) \, \varphi(\yy)\, e^{i(\xx-\yy)\cdot\pp/\hbar \, d\yy \, d\pp}.
\EE
We recall \cite{JungelBook,Tatarski83,Zachos05} that the inverse transform $\Op^{-1}$ coincides with the Wigner 
transform up to the identification of the operator $\Op(a)$ with its 
integral kernel $a$.
To this extent, the Wigner transform \eqref{WiT} and the Weyl transform \eqref{WeT} are the inverse of each other. 
\par
The definition of $g$ is completed by imposing the constraints \eqref{CEmom}.
In \eqref{gdef} the scalar function $A$ and the vector function $\BB = (B_1,B_2)$, 
both real valued,
are the Lagrange multipliers that provide the necessary degrees of freedom for $g$ to fulfil such constraints.
\subsection{Quantum Euler equations}

We now express the quantum Euler equations \eqref{EE1} in terms of the Lagrange multipliers. \\

\noindent  For the operator $\Lambda[\cE]$ we have 
$$
  \Lambda[\cE] g = \frac{i}{\hbar}\{ \cE, g\}_\# =  \frac{i}{\hbar}\{ \cE- \BB\cdot\pp - A, g\}_\#  + \frac{i}{\hbar}\{ \BB\cdot\pp + A, g\}_\#
$$ 
$$
 =  \frac{i}{\hbar}\{ \BB\cdot\pp + A, g\}_\# = \frac{i}{\hbar}\{ \BB\cdot\pp, g\}_\# + \Theta[A] g ,
$$
where we used the fact that $\{ \cE- \BB\cdot\pp - A, g\}_\#= 0$, since $g$ is a function of  (and therefore commutes with) $\cE- \BB\cdot\pp - A$ (see \eqref{gdef}).
Moreover (using the convention of summing over repeated indices $\BB\cdot\pp = B_j p_j$),
$$
\begin{aligned}
 \frac{i}{\hbar}\{ B_j\, p_j, g\}_\# &=  \frac{2i}{\hbar} \sum_{\text{odd}\ k} \hbar^k (B_jp_j) \#_k g 
 \\
&= \frac{2i}{\hbar} \, \sum_{k=0}^\infty \frac{\hbar^{2k+1}}{(2i)^{2k+1}} \sum_{\abs{\alpha}+\abs{\beta} = 2 \, k + 1} \frac{(-1)^\abs{\alpha}}{\alpha! \beta!}
   \nabla_x^\alpha B_j\nabla_p^\beta p_j \, \nabla_p^\alpha\nabla_x^\beta g .
\end{aligned}
$$
In the last expression, only the terms with $\beta=0$ and $\beta=1$ are nonzero in the factor $\nabla_p^\beta p_j$ and we obtain:
$$
\begin{aligned}
 \frac{i}{\hbar}\{ B_j\, p_j, g\}_\#  = &-\sum_{k=0}^\infty \frac{(-1)^{k} \hbar^{2k}}{4^k } \sum_{\abs{\alpha} = 2k+1} \frac{1}{\alpha!} \, p_j \nabla_x^\alpha B_j \nabla_p^\alpha g 
 \\
&+ \sum_{k=0}^\infty \frac{(-1)^{k} \hbar^{2k}}{4^k} \sum_{\abs{\alpha} = 2k} \frac{1}{\alpha!} \, \nabla_x^\alpha B_j \nabla_p^\alpha \frac{\pt g}{\pt x_j}.
\end{aligned}
$$
Comparing with \eqref{pseudoV} and \eqref{Thetaplus} yields
$$
\frac{i}{\hbar}\{ B_j\, p_j, g\}_\# = p_j\Theta[B_j] g + \Theta_+[B_j] \frac{\pt g}{\pt x_j}
$$
and, therefore, 
$$
\Lambda[\cE] g = p_j \Theta[B_j] g  + \Theta_+[B_j]\frac{\pt g}{\pt x_j} + \Theta[A]g \,.
$$
Then,
\BE
\label{Tg}
  Tg = \Lambda[\cE] g + \Theta[V]g = p_j \Theta[B_j] g  + \Theta_+[B_j]\frac{\pt g}{\pt x_j} + \Theta[A+V]g .
\EE
Here and in the following section we shall use the identities
\BE
\label{identities}
\begin{aligned}
&\bk{\Theta[V]w} = 0,&
&\bk{p_j\Theta[V]w} = \frac{\pt V}{\pt x_j}\,\bk{w},
\\
&\bk{\Theta_+[V]w} = V\bk{w},& \qquad
&\bk{p_j\Theta_+[V]w} = V\bk{p_jw}.
%\\
% & \bk{p_i \, p_j\Theta[V]w} 
% = \frac{\pt V}{\pt x_i}\,\bk{p_j \, w} + \frac{\pt V}{\pt x_j}\,\bk{p_i \, w}.
\end{aligned}
\EE
Thanks to \eqref{identities} we obtain from (\ref{Tg}):
\BE
\label{bkTg}
 \bk{T g}  = \bk{\Lambda[\cE] g + \Theta[V]g} = \frac{\pt }{\pt x_j} (B_j \bk{g}) =  \frac{\pt }{\pt x_j} (B_j n_0)
\EE
and, for $i = 1,2$,
\BE
\begin{aligned}
\label{bkpTg}
   \bk{p_i T g}  &= \bk{p_i\Lambda[\cE] g + p_i\Theta[V]g} 
\\
   &= \frac{\pt B_j }{\pt x_i} \bk{p_j g} + \frac{\pt }{\pt x_j} (B_j \bk{p_i g}) + \frac{\pt  (A+V)}{\pt x_i} \bk{g}
  \\
   &= \frac{\pt B_j }{\pt x_i} n_j + \frac{\pt }{\pt x_j} (B_j n_i) + \frac{\pt  (A+V)}{\pt x_i} n_0.
\end{aligned}
\EE
Hence, Eq.\ \eqref{EE1} reads as follows:
\BE
\label{QEE}
\left\{
\begin{aligned}
   &\pt_t n_0 + \frac{\pt (B_j n_0)}{\pt x_j} = 0
\\
   &\pt_t n_i + \frac{\pt B_j }{\pt x_i} n_j + \frac{\pt (B_j n_i)}{\pt x_j}   + \frac{\pt (A+V) }{\pt x_i} n_0 = 0,
   \qquad i = 1,2.
\end{aligned}
\right.
\EE
As discussed in Section \ref{SecCE}, equations \eqref{QEE} are the quantum Euler equations for electrons in graphene. 
The form of Eq.\ \eqref{QEE} is similar to that of the quantum Euler equations found in other contexts \cite{JSP12,DGM2007}.
%The semiclassical expansion such equations will be studied in 
%
%
\subsection{Quantum Navier-Stokes equations}
The next order in the Chapman-Enskog expansion, $\cO(\tau)$, provides the quantum Navier-Stokes equations which include viscosity terms. 
From equation (\ref{NS1}) we see that these terms are given by $\tau\bk{\psi_s TTg} + \tau \bk{\psi_s T \pt_t^0 g}$, for $s = 0,1,2$. In order to make these terms more explicit, we first write expression \eqref{Tg} in the more compact form
$$
  Tg = p_j\Theta[B_j] g  + \Theta_+[B_j]\frac{\pt g}{\pt x_j} + \Theta[A^*] g,
$$
where $A^* = A+V$.
Then, 
$$
TTg = \big(\Lambda[\cE] + \Theta[V]\big)\Big(  p_j\Theta[B_j] g  + \Theta_+[B_j]\frac{\pt g}{\pt x_j} + \Theta[A^*] g \Big) .
$$
By using \eqref{identities}, we have
\BE
\label{bkTTg}
\bk{TTg} = \bk{\Lambda[\cE]\Big(  p_j\Theta[B_j] g  + \Theta_+[B_j]\frac{\pt g}{\pt x_j} + \Theta[A^*] g \Big)} 
\EE
and
\BE
\label{bkpTTg}
\begin{aligned}
\bk{p_iTTg} =&\ \bk{\Lambda[\cE]\Big(  p_ip_j\Theta[B_j] g  + p_i\Theta_+[B_j]\frac{\pt g}{\pt x_j} + p_i \Theta[A^*] g \Big)} 
\\
 &+ \frac{\pt V}{\pt x_i}  \frac{\pt (B_j n_0)}{\pt x_j} .
\end{aligned}
\EE
Unfortunately, no simplifying identities like \eqref{identities} are available for $\Lambda[\cE]$.
Moreover, the non-polynomial form of the energy band $\cE$ makes the Navier-Stokes terms \eqref{bkTTg} and \eqref{bkpTTg} 
much less treatable than the corresponding terms for quadratic bands \cite{BM10}. 
\par
Turning to the expression $\bk{\psi_s T \pt_t^0 g}$, we recall that $\pt_t^0$ means that the time-derivatives of the moments $n_s$ are 
approximated by the Euler equations \eqref{QEE}. The term $\bk{\psi_s T \pt_t^0 g}$ cannot be made more explicit, given the lack of a general explicit form for the dependence of the Lagrange multipliers on the moments $n_s$.
This can only be done in the semiclassical approximation that will be discussed in Section \ref{SecSCL}.
For the moment, we shall just improve a little bit the expression by using
Eqs.\ \eqref{bkTg} and \eqref{bkpTg}:
\BE
\bk{T \pt_t^0 g} = \pt_t^0 \bk{Tg} = 
\pt_t^0 \left(\frac{\pt (B_j n_0)}{\pt x_j}\right) 
\EE
and, for $i = 1,2$, 
\BE
\bk{p_i T \pt_t^0 g} = 
\pt_t^0 \left(\frac{\pt B_j }{\pt x_i} n_j + \frac{\pt (B_j n_i)}{\pt x_j}   + \frac{\pt A^* }{\pt x_i} n_0\right).
\EE
\par
In conclusion, the quantum Navier-Stokes equations that we obtain are the following:
\begin{multline}
\label{QNS0}
  \pt_t n_0 + \frac{\pt (B_j n_0)}{\pt x_j} = 
 \tau \pt_t^0 \Big(\frac{\pt (B_j n_0)}{\pt x_j}\Big)
\\
  + \tau \bk{\Lambda[\cE]\Big(  p_j\Theta[B_j] g  + \Theta_+[B_j]\frac{\pt g}{\pt x_j} + \Theta[A^*] g \Big)} 
\end{multline}
and, for $i = 1,2$,
\begin{multline}
\label{QNS1}
  \pt_t n_i + \frac{\pt B_j }{\pt x_i} n_j + \frac{\pt (B_j n_i)}{\pt x_j}   + \frac{\pt A^* }{\pt x_i} n_0 = 
\\
 \tau \pt_t^0\Big(\frac{\pt B_j }{\pt x_i} n_j + \frac{\pt (B_j n_i)}{\pt x_j}   + \frac{\pt A^* }{\pt x_i} n_0\Big) 
+\tau\frac{\pt V}{\pt x_i}  \frac{\pt (B_j n_0)}{\pt x_j} 
\\
  + \tau \bk{\Lambda[\cE]\Big(  p_i p_j\Theta[B_j] g  + p_i\Theta_+[B_j]\frac{\pt g}{\pt x_j} + p_i \Theta[A^*] g \Big)}.
\end{multline}
Here, $A^* = A+V$ and $\pt_t^0$ means that every time-derivative of the
moments $n_0$, $n_1$ and $n_2$ must be computed by using
\BE
\begin{aligned}
&\pt_t^0 n_0 =  - \frac{\pt (B_j n_0)}{\pt x_j} ,
\\
&\pt_t^0 n_i = - \frac{\pt B_j }{\pt x_i} n_j -\frac{\pt (B_j n_i)}{\pt x_j}   - \frac{\pt (A+V) }{\pt x_i} n_0,
  \qquad i = 1,2.
\end{aligned}
\EE
By tedious calculations, reported in the Appendix, equations (\ref{QNS0}) and (\ref{QNS1}) can be expanded as:
\BE
\begin{aligned}
&\pt_t n_0 + \frac{\pt (B_j n_0)}{\pt x_j} 
= \tau \left\{ \frac{\delta }{\delta n_k} \left(\frac{\partial (B_j \, n)}{\partial x_j}\right)\pt_t^0 n_k
 \right. 
\\
&\quad -\sum_{k=0}^{\infty} \sum_{l=0}^{\infty} \gamma_{kl} \sum_{|\alpha|=2k+1}\sum_{|\beta|=2l+1} \frac{1}{\alpha! \, \beta!}
\left[ \nabla_{\xx}^\alpha((\nabla_{\xx}^\beta B_j) \mathcal{H}_j^{\alpha \beta}(x,p)) \right.
\\
&\quad \left. \left. - \nabla_{\xx}^\alpha \left( (\nabla_{\xx}^\beta B_j) \frac{\partial}{\partial x_j}\mathcal{G}^{\alpha \beta}(x,p)\right)+\nabla_{\xx}^\alpha \left((\nabla_{\xx}^\beta A^*) \mathcal{G}^{\alpha \beta}(x,p)\right) \right]\right\}
\label{NavSto1} 
\end{aligned}
\EE
\BE
\begin{aligned}
&\pt_t n_i + \frac{\pt B_j }{\pt x_i} n_j + \frac{\pt (B_j n_i)}{\pt x_j}   + \frac{\pt A^* }{\pt x_i} n_0 
\\
&\quad =  \tau \left\{\left[\frac{\delta }{\delta n_k} \left(\frac{\pt B_j }{\pt x_i} n_j\right) + \frac{\delta }{\delta n_k} \left(\frac{\pt }{\pt x_j} (B_j n_i)\right)  \frac{\delta }{\delta n_k} \left(\frac{\pt  A}{\pt x_i} n_0 \right) \right.\right. 
\\
& \left. \qquad +\frac{\pt  V}{\pt x_i} \delta_{k0}  \right] \, \pt_t^0 n_k +\frac{\pt V}{\pt x_i}  \frac{\pt (B_j n_0)}{\pt x_j}
\\
&\qquad-\sum_{k=0}^{\infty} \sum_{l=0}^{\infty} \gamma_{kl} \sum_{|\alpha|=2k+1}\sum_{|\beta|=2l+1} \frac{1}{\alpha! \, \beta!}
\left[ \nabla_{\xx}^\alpha((\nabla_{\xx}^\beta B_j) \mathcal{K}_{ij}^{\alpha \beta}(x,p))- \right. 
\\
&\quad \left. \left. - \nabla_{\xx}^\alpha \left( (\nabla_{\xx}^\beta B_j) \frac{\partial}{\partial x_j}\mathcal{H}_i^{\alpha \beta}(x,p)\right)+\nabla_{\xx}^\alpha \left((\nabla_{\xx}^\beta A^*) \mathcal{H}_i^{\alpha \beta}(x,p)\right) \right]\right\} ,
\end{aligned}
\label{NavSto2}
\EE
where we have introduced the quantities
\BE
\begin{aligned}
&\mathcal{G}^{\alpha \beta}(x,p)= \bk{(\nabla_{\pp}^\alpha \cE) \nabla_{\pp}^\beta g}, 
\\
&\mathcal{H}_j^{\alpha \beta}(x,p) = \bk{p_j(\nabla_{\pp}^\alpha \cE) \nabla_{\pp}^\beta g)},
\\
&\mathcal{K}_{ij}^{\alpha \beta}(x,p) = \bk{p_i \, p_j \, (\nabla_{\pp}^\alpha \cE) \nabla_{\pp}^\beta g)}.
\end{aligned}
\EE
No simpler form of these equations can be obtained in general; we shall introduce their semiclassical approximation in the next section.

\section{Semiclassical expansion}
\label{SecSCL}
In this section we introduce the semiclassical approximation, by expanding all quantities and equations to second order in $\hbar$. To this aim, we introduce the phase space function $h(\xx,\pp)$ given by
\BE
\label{hdef}
 h := -\cE + \BB\cdot\pp + A.
\EE
where the energy band $\cE=\cE(\pp)$ is given by (\ref{Eksmooth}); $A=A(\xx)$ and $\BB=\BB(\xx)$ are the Lagrange multipliers. For the sake of simplicity, let us assume that the Fermi-Dirac distribution can be
approximated with the Maxwell-Boltzmann one. 
Eq. \eqref{gdef} then becomes
\BE
\label{qexp}
    g = \Op^{-1} \left( \e^{\Op(h)} \right),
\EE
which is the so-called {\em quantum exponential} of $h$ \cite{JSP12,DGM2007,DMR05,DR03}.
The quantum exponential admits the semiclassical expansion
\BE
  g = g^{(0)} + \hbar^2 g^{(2)} + \cO(\hbar^4), \label{gexp}
\EE
where
\BE
  g^{(0)} = \e^h
\EE
is the classical exponential and \cite{JSP12,DGM2007,DR03,JungelBook}
\BE
\label{g2def}
\begin{aligned}
g^{(2)} = \frac{1}{8} \e^h\Big\{ &-\frac{\pt^2 h}{\pt x_i\pt x_j}\frac{\pt^2 h}{\pt p_i\pt p_j} + \frac{\pt^2 h}{\pt x_i\pt p_j}\frac{\pt^2 h}{\pt p_i\pt x_j}
-\frac{1}{3}\frac{\pt^2 h}{\pt x_i\pt x_j}\frac{\pt h}{\pt p_i}\frac{\pt h}{\pt p_j}
\\
&+\frac{2}{3}\frac{\pt^2 h}{\pt x_i\pt p_j}\frac{\pt h}{\pt p_i}\frac{\pt h}{\pt x_j}
-\frac{1}{3}\frac{\pt^2 h}{\pt p_i\pt p_j}\frac{\pt h}{\pt x_i}\frac{\pt h}{\pt x_j}
\Big\},
\end{aligned}
\EE
where summation convention over repeated indices is adopted. 
In our case, recalling \eqref{velocity}, we have
$$
\frac{\pt h}{\pt x_i} = 
\frac{\pt (\BB\cdot\pp + A)}{\pt x_i},
\qquad \frac{\pt h}{\pt p_i} = B_i - v_i ,
$$ $$
\frac{\pt^2 h}{\pt x_i \pt x_j} = 
\frac{\pt^2 (\BB\cdot\pp + A)}{\pt x_i \pt x_j}
 ,
\qquad
\frac{\pt^2 h}{\pt p_i \pt p_j} = - \frac{\pt v_i}{\pt p_j},
\qquad
\frac{\pt^2 h}{\pt x_i \pt p_j} = \frac{\pt B_j}{\pt x_i}.
$$
Substituting into \eqref{g2def} yields
\BE
\label{g2explicit}
\begin{aligned}
g^{(2)} =
 \frac{1}{8} \e^h\Big\{ &\frac{\pt^2 (\BB\cdot\pp + A)}{\pt x_i \pt x_j} \frac{\pt v_i}{\pt p_j}
 +\frac{\pt B_j}{\pt x_i} \frac{\pt B_i}{\pt x_j}
\\
 &-\frac{1}{3} \frac{\pt^2(\BB\cdot\pp + A)}{\pt x_i \pt x_j}
  (B_i-v_i)(B_j-v_j)
\\
 &+\frac{2}{3} \frac{\pt B_j}{\pt x_i} \frac{\pt (\BB\cdot\pp + A)}{\pt x_j} (B_i-v_i) 
\\
 &+ \frac{1}{3}\frac{\pt (\BB\cdot\pp + A)}{\pt x_i}
 \frac{\pt (\BB\cdot\pp + A)}{\pt x_j} \frac{\pt v_i}{\pt p_j}
\Big\}.
\end{aligned}
\EE
\subsection{Semiclassical expansion of the constraints}
In order to proceed with the semiclassical approximation, we need 
 to expand the constraints \eqref{CEmom} up to order $\hbar^2$, i.e.
\BE
\label{SCconstraint}
  \bk{g^{(0)}} + \hbar^2 \bk{g^{(2)}} = n_0,
  \qquad
  \bk{\pp g^{(0)}} + \hbar^2 \bk{\pp g^{(2)}} = n_\pp.
\EE
We begin with the leading order, $g^{(0)} = \e^h$. 
By using the polar representations 
$$
  \pp = \rho (\cos\theta, \sin\theta),
  \qquad
 \BB = b(\cos\theta_B, \sin\theta_B),
$$
we obtain
$$
\bk{e^h}= \int_{\mR^2} \e^{-\cE + \BB\cdot\pp + A } d\pp =
 \e^A\int_0^\infty \rho\, \e^{-c\sqrt{\rho^2+\alpha^2}}\int_{-\pi}^{\pi} \e^{b\rho\cos(\theta-\theta_B)} d\theta d\rho .
$$ $$
= 
 \e^A\int_0^\infty \rho\, \e^{-c\sqrt{\rho^2+\alpha^2}}\int_{-\pi}^{\pi} \e^{b\rho\cos(\theta)} d\theta d\rho
 = 2\pi \e^A\int_0^\infty \rho\, \e^{-c\sqrt{\rho^2+\alpha^2}}
 I_0(b\rho)\,d\rho
$$ $$
= 2\pi\e^A \int_\alpha^\infty s \, \e^{-cs} I_0(b\sqrt{s^2- \alpha^2})\,ds
= 2\pi\e^A \alpha^2 \int_1^\infty  t\,\e^{-c\alpha t} I_0(b\alpha\sqrt{t^2- 1})\,dt
$$ $$
= -2\pi\e^A \frac{\pt}{\pt c} \left(
\alpha \int_1^\infty \e^{-c\alpha t} I_0(b\alpha\sqrt{t^2- 1})\,dt.
\right)
$$
Since
$$
\alpha \int_1^\infty \e^{-c\alpha t} I_0(b\alpha\sqrt{t^2- 1})\,dt = 
\frac{\e^{-\alpha\sqrt{c^2-b^2}}}{\sqrt{c^2-b^2}}
$$
(see e.g.\ Ref.\ \cite{GR2000}), we obtain
\BE
\label{bkeh}
\bk{e^h} = -2\pi\alpha\e^A\frac{\pt}{\pt c} \left( \frac{\e^{-\alpha\sqrt{c^2-b^2}}}{\sqrt{c^2-b^2}} \right)
= \frac{2\pi c\,\e^A\e^{-\alpha\sqrt{c^2-b^2}} (\alpha\sqrt{c^2-b^2}+1)}{(c^2-b^2)^{3/2}}
\EE
Now, 
$$
  \nabla_\pp \e^h = -\vv \e^h + \BB \e^h
$$
and so we have
$$
\bk{\vv \e^h} = \BB \bk{\e^h}.
$$
Moreover,
\BE
\label{bkpeh}
\begin{aligned}
\bk{\pp \e^h} &= \bk{\nabla_{\BB}\, \e^h} = \nabla_{\BB} \bk{\e^h} 
= \nabla_{\BB} \left(\frac{2\pi c\,\e^A\e^{-\alpha\sqrt{c^2-b^2}} (\alpha\sqrt{c^2-b^2}+1)}{(c^2-b^2)^{3/2}}\right)
\\
&= \frac{2\pi c\,\e^A \e^{-\alpha\sqrt{c^2-b^2}} \big( \alpha^2(c^2-b^2) +3\alpha\sqrt{c^2-b^2}+ 3\big)}{(c^2-b^2)^{5/2}} \BB
\\
&=  \left( \frac{3}{c^2-b^2} + \frac{\alpha^2}{\alpha\sqrt{c^2-b^2}+1} \right)\bk{\e^h} \BB .
\end{aligned}
\EE
It is convenient to define $\jj$ such that
\BE
\label{jdef}
  n_\pp = n_0\jj
\EE
and to introduce the semiclassical current $\uu$ as
\BE
n_0\uu := \bk{\vv \e^h}.
\EE
Hence, from the preceding computations, we see that the following relations between the Lagrange multipliers $A$ and $\BB$ 
and the hydrodynamic moments $n_0$, $\uu$ and $n_\pp$
hold true at leading order:
\BE
\label{LO}
\begin{aligned}
&n_0 = \frac{2\pi c\,\e^A\e^{-\alpha\sqrt{c^2-u^2}} (\alpha\sqrt{c^2-u^2}+1)}{(c^2-u^2)^{3/2}} ,
\\
&\uu = \BB ,
\\
&\jj = \left( \frac{3}{c^2-u^2} + \frac{\alpha^2}{\alpha\sqrt{c^2-u^2}+1} \right) \uu ,
\end{aligned}
\EE
where, as usual, $u := \abs{\uu}$.
\begin{rem}
Since $\e^h/n_0$ is a non-constant probability distribution in $\pp$, and since $\abs{\vv} \leq c$ (for $\alpha \geq 0$), then from Jensen's inequality we have
$$
  \abs{\uu} = \frac{1}{n_0} \abs{ \bk{\vv \e^h} }
  < \frac{1}{n_0} \bk{\abs{\vv} \e^h} \leq c,
$$
with the strict inequality sign,
which ensures that the consistency relation $u < c$ is fulfilled (also for $\alpha = 0$).
\end{rem}
\begin{rem}
For $\alpha = 0$ we obtain the relations 
\BE
n_0 = \frac{2\pi c\, \e^A}{(c^2-u^2)^{3/2}} ,
\qquad
\uu = \BB ,
\qquad
\jj = \frac{3\uu}{c^2 - u^2}
\EE
which are known to hold true for the conical band (with Maxwell-Boltzmann statistics), see e.g.\ Ref.\ \cite{Narozhny19}.
\end{rem}
To write down the constraints \eqref{SCconstraint} at second order, we need to compute $\bk{g^{(2)}}$
and $\bk{\pp g^{(2)}}$, i.e.\ $\bk{\psi_s g^{(2)}}$ for $s = 0,1,2$.
From \eqref{g2explicit} we have: 
\BE
\label{g2moment0}
\begin{aligned}
\bk{\psi_s g^{(2)}} =
 \frac{1}{8} \Big\{ 
 &\bk{\psi_s\frac{\pt^2 (\BB\cdot\pp + A)}{\pt x_i \pt x_j} \frac{\pt v_i}{\pt p_j}\, \e^h }
 +\frac{\pt B_j}{\pt x_i} \frac{\pt B_i}{\pt x_j}\,n_s
\\
 &-\frac{1}{3} \bk{\psi_s \frac{\pt^2(\BB\cdot\pp + A)}{\pt x_i \pt x_j}
  (B_i-v_i)(B_j-v_j) \e^h }
\\
 &+\frac{2}{3} \frac{\pt B_j}{\pt x_i} \bk{ \psi_s \frac{\pt (\BB\cdot\pp + A)}{\pt x_j} (B_i-v_i)\,\e^h }
\\
 &+ \frac{1}{3}\bk{ \psi_s \frac{\pt (\BB\cdot\pp + A)}{\pt x_i}
 \frac{\pt (\BB\cdot\pp + A)}{\pt x_j} \frac{\pt v_i}{\pt p_j} \,\e^h }
 \Big\} 
\end{aligned}
\EE
for $s = 0,1,2$.
\par
The moments (i.e.\ the expressions in angular brackets) in Eq.\ \eqref{g2moment0},
can be (in principle) computed explicitly as functions of $n_0$ and $\uu$ by using techniques similar to those employed 
to obtain the relations \eqref{LO}, but the resulting expressions are much more cumbersome.
\par
Not that all the moments appearing in \eqref{g2moment0} are finite also for $\alpha = 0$
(because the only singularity comes from $\pt v_i / \pt p_j$, which behaves like $p^{-1}$, and thus  integrable in dimension two).
This implies that the semiclassical expansion of the Lagrange multipliers has no singularities, at least to second order.
On the other hand, the quantum Euler equations \eqref{QEE} have been proven to depend only on the Lagrange multipliers.
Hence, we obtain the remarkable result that, in our approach, the second-order semiclassical Euler equations contain no singularities, 
even without band regularization.
\subsection{Inversion of the constraint system}
Equations \eqref{LO} and \eqref{g2moment0} are the expressions of the hydrodynamic moments 
as functions of the Lagrange multipliers to second order (first order in $\hbar^2$).
However, in order to get explicit semiclassical Navier-Stokes equations, it is necessary to express 
the Lagrange multipliers in terms of the hydrodynamic moments in the quantum 
Navier-Stokes equations \eqref{QNS0}-\eqref{QNS1}.
Then, our next task is to invert the $\cO(\hbar^2)$ constraint system  \eqref{SCconstraint}. 
\par
To better understand this point, let us schematically indicate by $\vecA = (A,\BB)$ the vector of Lagrange multipliers, 
and by $\vecN = (n_0,\jj)$ the vector of hydrodynamic moments.
%and by $\vecpsi = (\psi_1,\psi_2,\psi_2)$ the vector of hydrodynamic monomials.
In the previous section, we have expressed $\vecN$ as a function of $\vecA$ to first order in $ \hbar^2$:
\BE
\label{NbyA}
  \vecN = \FF^{(0)}(\vecA) + \hbar^2 \FF^{(2)}(\vecA) + \cO(\hbar^4),
\EE
where $\FF^{(0)}(\vecA)$ and  $\FF^{(2)}(\vecA)$ are given by Eqs. \eqref{LO} and \eqref{g2moment0}, respectively. To invert these relations at first order (in $\hbar^2$) means writing
\BE
\label{AbyN}
  \vecA = \GG^{(0)}(\vecN) + \hbar^2\GG^{(2)}(\vecN) + \cO(\hbar^4).
\EE
To determine the functions $\GG^{(0)}$ and $\GG^{(2)}$, we insert \eqref{AbyN} into \eqref{NbyA}, which yields (by Taylor expansion)
$$
 \vecN = \FF^{(0)} \big(\GG^{(0)}(\vecN)\big) + \hbar^2 \frac{\pt\FF^{(0)}}{\pt \vecA}\big(\GG^{(0)}(\vecN)\big) \GG^{(2)}(\vecN) 
 + \hbar^2 \FF^{(2)}\big(\GG^{(0)}(\vecN)\big) + \cO(\hbar^4).
$$
At leading order we obtain, of course, that $\GG^{(0)} = \big( \FF^{(0)} \big)^{-1}$.
The leading-order expression of the Lagrange multipliers as a function of $\vecN$,
$$
  \vecA^{(0)} = \GG^{(0)}(\vecN) = \big( \FF^{(0)} \big)^{-1}(\vecN)
$$
corresponds to the inversion of \eqref{LO}.
The next-order inversion formula is the linear system
$$
 \frac{\pt\FF^{(0)}}{\pt \vecA} (\vecA^{(0)})\, \vecA^{(2)} =
 - \FF^{(2)}(\vecA^{(0)}) 
$$
in the unknown
$$
  \vecA^{(2)} := \GG^{(2)}(\vecN)
$$ 
(see also \cite{JSP12}).
We remark that $\vecA^{(0)}$ and $\vecA^{(1)}$ depend on $\vecN$.
\par
At leading order, the inversion of system \eqref{LO} leads to $\vecA^{(0)} = (A^{(0)}, \BB^{(0)})$, where
\BE
\label{invLO}
\begin{aligned}
 &A^{(0)} = a(n_0,u) = \log\left( \frac{(c^2-u^2)^{3/2} \, n_0} {2\pi c\,\e^{-\alpha\sqrt{c^2-u^2}} (\alpha\sqrt{c^2-u^2}+1)} \right),
 \\
 &\BB^{(0)} = \uu ,
\end{aligned}
\EE
where the link with $\jj$ is furnished by the leading-order relation
\BE
\label{JLO}
\jj = \left( \frac{3}{c^2-u^2} + \frac{\alpha^2}{\alpha\sqrt{c^2-u^2}+1} \right) \uu.
\EE
\par
At second order, as we have just seen, $\vecA^{(2)} = (A^{(2)}, \BB^{(2)})$ is the solution of the linear system
$$
\frac{\pt\FF^{(0)}}{\pt \vecA} (\vecA^{(0)})\, \vecA^{(2)} =  - \FF^{(2)}(\vecA^{(0)}),
$$
where 
$$
\frac{\pt\FF^{(0)}}{\pt \vecA} (\vecA^{(0)})
$$ 
is the Jacobian matrix of the mapping $\eqref{LO}$ evaluated at 
$A = A^{(0)}$, $\BB = \BB^{(0)}$ (given by \eqref{invLO}), and $\FF^{(2)}(\vecA^{(0)})$ is given by 
\eqref{g2moment0}, also evaluated at $A = A^{(0)}$, $\BB = \BB^{(0)}$.
More explicitly, using \eqref{bkeh} and \eqref{bkpeh},
\BE
 \frac{\pt\FF^{(0)}}{\pt \vecA} (\vecA^{(0)}) = 
 \begin{pmatrix}
 \frac{\pt \langle\e^h\rangle}{\pt A} &  \frac{\pt \langle\e^h\rangle}{\pt \BB}
 \\[4pt]
 \frac{\pt \frac{1}{n_0}\langle\pp\e^h\rangle}{\pt A} &  \frac{\pt \frac{1}{n_0}\langle\pp\e^h\rangle}{\pt \BB}
 \end{pmatrix}_{
 \big \vert A = A^{(0)}, \BB = \BB^{(0)} }
=  \begin{pmatrix}
n_0 & n_0\jj
\\
\boldsymbol{0} & M(n_0,\uu)
\end{pmatrix}
\EE
where $M(n_0,\uu)$ is the $2\times 2$ matrix
\begin{multline}
M(n_0,\uu) = \frac{\alpha}{\sqrt{c^2-u^2} (\alpha \sqrt{c^2-u^2}+ 1)} \BB\otimes\BB
\\
+ \left(\frac{3}{c^2-u^2}+ \frac{\alpha^2}{\alpha\sqrt{c^2-u^2}+1} \right) I.
\end{multline}
Moreover,
$$
\FF^{(2)}(\vecA^{(0)}) = \begin{pmatrix}
f_1^{(2)}(n_0,\uu)
\\
f_\pp^{(2)}(n_0,\uu)
\end{pmatrix}
$$
where $f_1^{(2)}(n_0,\uu)$ and $f_\pp^{(2)}(n_0,\uu)$ are given by the right-hand side of \eqref{g2moment0}, 
for $s = 0,1,2$,
with $A$, $\BB$ substituted by their leading-order expressions \eqref{invLO}, and 
$\jj$ is still given by the leading-order relation \eqref{JLO}.
Hence, the second-order corrections to the relations \eqref{invLO} are obtained by solving the linear system
\BE
\begin{pmatrix}
n_0 & n_0 \jj
\\
\boldsymbol{0} & M(n_0,\uu)
\end{pmatrix}
\begin{pmatrix}
A^{(2)} \\ \BB^{(2)}
\end{pmatrix}
=
- \begin{pmatrix}
f_1^{(2)}(n_0,\uu)
\\
f_\pp^{(2)}(n_0,\uu)
\end{pmatrix} \label{invL2}
\EE
(together with the relation \eqref{JLO} between $\jj$ and $\uu$).
\subsection{Semiclassical expansion of the quantum Navier-Stokes equations}
We are now able to expand the quantum Navier-Stokes equations \eqref{QNS0}-\eqref{QNS1} to
order $\hbar^2$.
To this aim, we recall from Sect.\ \ref{S2} that the pseudo-differential operator
admit the expansions:
\BEA
&& \Lambda[\cE] \approx \Lambda^{(0)}[\cE] + \hbar^2\Lambda^{(2)}[\cE] 
= \vv \cdot \nabla_\xx -  \frac{\hbar^2}{4}\sum_{\abs{\alpha} = 3}
\frac{1}{\alpha!} \nabla_\pp^\alpha \cE \cdot \nabla_\xx^\alpha ,
\label{Lambdaexp} 
\\
&& \Theta[V] \approx \Theta^{(0)}[V] + \hbar^2\Theta^{(2)}[V]
= - \nabla_\xx V \cdot \nabla_\pp + \frac{\hbar^2}{4}\sum_{\abs{\alpha} = 3}
\frac{1}{\alpha!} \nabla_\xx^\alpha V \cdot \nabla_\pp^\alpha , \phantom{++} 
\label{Thetaexp} 
\\
&& \Theta_+[V] \approx \Theta_+^{(0)}[V] + \hbar^2\Theta_+^{(2)}[V]
= V - \frac{\hbar^2}{4}\sum_{\abs{\alpha} = 2}
\frac{1}{\alpha!} \nabla_\xx^\alpha V \cdot \nabla_\pp^\alpha. 
\label{Thetaplusexp}
\EEA
These expansions, together with the expansions of the equilibrium and of the Lagrange multipliers studied in the
previous section, will now be used in equations \eqref{QNS0}-\eqref{QNS1} to obtain the $\cO(\hbar^2)$ semiclassical model.
For the sake of clarity, let us write the $\cO(\hbar^2)$ model in this way
\BE
\pt_t n_i + E_k^{(0)} + \hbar^2E_i^{(2)} = \tau S_i^{(0)} + \tau\hbar^2S_i^{(2)}, \qquad i = 0,1,2
\EE
and compute the expressions $E_i^{(m)}$ and $S_i^{(m)}$ one by one.
Note that $E_i^{(0)}$ and $E_i^{(2)}$ provide the leading-order and second-order approximations of the quantum Euler equations, 
while $S_i^{(0)}$ and $S_i^{(2)}$ are the analogous approximation of the ``viscous'' terms.
\par
From Eqs.\ \eqref{QNS0} and the leading-order expansion \eqref{invLO} of the Lagrange multipliers,  we immediately obtain for the Euler terms $E_i^{(0)}$:
\BE
\label{E0}
\begin{aligned}
  &E_0^{(0)} = \frac{\pt (n_0 u_j )}{\pt x_j} , %= \DIV (n_0\uu)
\\
  &E_i^{(0)} =  n_j \frac{\pt u_j }{\pt x_i} + \frac{\pt (n_i u_j )}{\pt x_j} + n_0 \frac{\pt (a + V)}{\pt x_i}, \qquad  i = 1,2
\end{aligned}
\EE
where $a = A^{(0)}$ is explicitly given by \eqref{invLO} as a function of $n_0$ and $u = \abs{\uu}$.
The $\hbar^2$ quantum corrections
\BE
\label{E2}
\begin{aligned}
  &E_0^{(2)} = \frac{\pt (n_0 B_j^{(2)} )}{\pt x_j} , 
\\
  &E_i^{(2)} =  n_j \frac{\pt B_j^{(2)}}{\pt x_i} + \frac{\pt (n_i B_j^{(2)} )}{\pt x_j} + n_0 \frac{\pt A^{(2)}}{\pt x_i}, \qquad  i = 1,2,
\end{aligned}
\EE
have a similar structure in terms of the second-order Lagrange multipliers $A^{(2)}$ and $B_j^{(2)}$, but the dependence of the latter 
on the hydrodynamic unknowns $n_i$ is complicated and it is not explicitly written down here: it can be deduced starting 
from Eq.\ \eqref{invL2}. 
\par
Let us now turn to the viscosity terms, starting from the leading-order.
From \eqref{QNS0}, \eqref{invLO} and \eqref{Lambdaexp}-\eqref{Thetaplusexp} we obtain
\begin{multline*}
  S_0^{(0)}  = \pt_t^0  \,\frac{\pt (n_0 u_j)}{\pt x_j}
 \\
  + \frac{\pt}{\pt x_k} \left[ -\frac{\pt u_j}{\pt x_l}   \bk{v_k  p_j \frac{\pt g^{(0)} }{\pt p_l}}  + u_j \frac{\pt  \bk{v_k g^{(0)}}}{\pt x_j} 
  - \frac{\pt (a+V)}{\pt x_l} \bk{ v_k \frac{\pt g^{(0)} }{\pt p_l}} \right],
\end{multline*}
where $g^{(0)}$ is the local equilibrium at leading order, i.e.
$$
   g^{(0)} = \exp\left( -\cE + \uu\cdot\pp + a \right).
$$
Keeping in mind the relationships 
\eqref{jdef} and \eqref{JLO} between $\uu$, $\jj$, $n_0$ and $n_\pp$, we have
$$
\frac{\pt (n_0u_j)}{\pt n_0} = u_j + n_0 \frac{\pt u_j}{\pt n_0} 
$$
and, for $i = 1,2$,
$$
\frac{\pt (n_0u_j)}{\pt n_i} = \frac{\pt (n_0u_j)}{\pt j_k}\frac{\pt j_k}{\pt n_i} = n_0 \frac{\pt u_j}{\pt j_k} \frac{1}{n_0} = \frac{\pt u_j}{\pt j_k},
$$
where $\pt u_j/\pt j_k$ is to be computed from Eq.\ \eqref{JLO}.
Note that here, at variance with Section \ref{S3.1}, we are using ordinary derivatives with respect to the moments $n_i$: this is because,
in the semiclassical expansion, the Lagrange multipliers depend locally on the moments.
Therefore, also recalling that $\bk{v_k g^{(0)}} = n_0 u_k$ and
$$
  \pt_t^0 n_i = -E^{(0)}_i,   \qquad i = 0,1,2,
$$
we rewrite the expression for $S_0^{(0)} $ as follows:
\begin{multline}
  S_0^{(0)}  =  -\frac{\pt }{\pt x_j} \left[ \left( u_j + n_0 \frac{\pt u_j}{\pt n_0} \right) E^{(0)}_0  +  \frac{\pt u_j}{\pt j_k} E^{(0)}_k\right]
 \\
  +\frac{\pt}{\pt x_k} \left( \frac{\pt u_j}{\pt x_l}   \bk{\frac{\pt  (v_k  p_j)}{\pt p_l} g^{(0)} }  + u_j \frac{\pt  (n_0 u_k)}{\pt x_j} 
  +  \frac{\pt (a+V)}{\pt x_l} \bk{ \frac{\pt  v_k}{\pt p_l} g^{(0)} } \right),
\end{multline}
where also integrations by parts have been performed.
By analogous procedures, we first obtain from Eq.\ \eqref{QNS1}:
\begin{multline*}
  S_i^{(0)} = \pt_t^0 \Big(\frac{\pt u_j }{\pt x_i} n_j + \frac{\pt (n_i u_j)}{\pt x_j}   + \frac{\pt (a+V) }{\pt x_i} n_0\Big)
+\frac{\pt V}{\pt x_i}  \frac{\pt (n_0 u_j)}{\pt x_j} 
\\
  +\frac{\pt}{\pt x_k} \left[ -\frac{\pt u_j}{\pt x_l}   \bk{v_k p_i  p_j \frac{\pt g^{(0)} }{\pt p_l}}  + u_j \frac{\pt  \bk{v_k p_i g^{(0)}}}{\pt x_j} 
  - \frac{\pt (a+V)}{\pt x_l} \bk{ v_k p_i \frac{\pt g^{(0)} }{\pt p_l}} \right]
\end{multline*}
and then 
\begin{multline}
  S_i^{(0)} = -\left[ n_j \frac{\pt}{\pt x_i} \frac{\pt u_j}{\pt n_0} + \frac{\pt}{\pt x_i}\left(n_i \frac{\pt u_j}{\pt n_0} \right)
  + n_0 \frac{\pt}{\pt x_i} \frac{\pt a}{\pt n_0} + \frac{\pt a}{\pt x_i} \right] E^{(0)}_0
\\  
  -\left[ n_k \frac{\pt}{\pt x_i} \frac{\pt u_j}{\pt n_k} + \frac{\pt u_j}{\pt x_k} 
  + \frac{\pt}{\pt x_j}\left(u_j \delta_{ik} + n_i \frac{\pt u_j}{\pt n_k} \right)
  + n_0 \frac{\pt}{\pt x_i} \frac{\pt a}{\pt n_k}\right] E^{(0)}_k 
\\
  +\frac{\pt}{\pt x_k} \left[ \frac{\pt u_j}{\pt x_l}   \bk{\frac{\pt(v_k p_i  p_j)}{\pt p_l} g^{(0)}}  + u_j \frac{\pt  \bk{v_k p_i g^{(0)}}}{\pt x_j} 
  + \frac{\pt (a+V)}{\pt x_l} \bk{ \frac{\pt (v_k p_i) }{\pt p_l} g^{(0)} } \right]
\end{multline}
(for $i = 1,2$), where the terms  
$$
-\frac{\pt V}{\pt x_i} E^{(0)}_0 \qquad \mbox{and} \qquad \frac{\pt V}{\pt x_i}  \frac{\pt (n_0 u_j)}{\pt x_j}
$$
have canceled.
Working out explicitly the last term, $S_i^{(2)}$, leads to very cumbersome expressions and therefore we limit ourselves to report its overall form, 
without entering into further details.
For $i = 0$ we have
\begin{multline}
  S_0^{(2)} =  \pt_t^0 \left(\frac{\pt (n_0 B_j^{(2)})}{\pt x_j}\right)
\\[4pt]
  +\!\!\!\!\sum_{\alpha+\beta+\gamma+\delta = 2} \bk{\Lambda^{(\alpha)}[\cE] \Big( \Theta^{(\beta)}[p_j B_j^{(\gamma)} + A^{(\gamma)}_*] g^{(\delta)} 
  + \Theta^{(\beta)}_+[B_j^{(\gamma)}]\frac{\pt g^{(\delta)}}{\pt x_j} \Big)}
\end{multline}
and, for $i = 1,2$,
\begin{multline}
  S_i^{(2)} =  \pt_t^0 \left(\frac{\pt B_j^{(2)} }{\pt x_i} n_j + \frac{\pt (n_i B_j^{(2)})}{\pt x_j}   + \frac{\pt A^{(2)}}{\pt x_i} n_0\right)
  +\frac{\pt V}{\pt x_i}  \frac{\pt (n_0 B_j^{(2)})}{\pt x_j} 
  \\[4pt]
  +\!\!\!\!\sum_{\alpha+\beta+\gamma+\delta = 2} \bk{\Lambda^{(\alpha)}[\cE] \Big( \Theta^{(\beta)}[p_i p_j B_j^{(\gamma)} + p_i A^{(\gamma)}_*] g^{(\delta)} 
  + \Theta^{(\beta)}_+[p_i B_j^{(\gamma)}]\frac{\pt g^{(\delta)}}{\pt x_j} \Big)}.
\end{multline}
\section{Conclusions}
In this work, we have obtained the hydrodynamic equations for a population of conduction electron carriers in graphene, by using a regularised form for the energy band and by assuming a isothermal regime. 
The macroscopic equation include the inviscid Euler equations and the viscosity terms. 
The underlying kinetic model is given by the Wigner-function approach, in which the effect of the periodic potential is accounted for by a pseudo-differential operator with the symbol given by the band shape. 
We use the Chapman-Enskog procedure accompanied by the Quantum Maximum Energy Principle, which provides the Wigner equilibrium function in terms of a set of Lagrange multipliers. 
The relationship between the hydrodynamic quantities and the Lagrange multipliers cannot be inverted in general, thus we obtain hydrodynamic equations which, in their general quantum form, are left implicit. 
Subsequently, we expand the hydrodynamic equations together with all relevant quantities to second order in $\hbar$, thus obtaining the semiclassical approximation for the inviscid and viscous hydrodynamic equations.
It is worth to remark that, in contrast to previous studies, the semiclassical Euler equations that we obtain by applying the QMEP are non-singular in the zero-gap limit $\alpha \to 0$.
\section*{Appendix}
We report here some of the details of the algebraic steps leading to the explicit Navier-Stokes equations in the form \eqref{NavSto1}) and \eqref{NavSto2} starting from \eqref{QNS0} and \eqref{QNS1}.
\par
We introduce the quantity
$$
\gamma_{kl}= \frac{(-1)^{k+l}\hbar^{2(k+l)}}{2^{2(k+l)}},
$$
so that the expressions involved in \eqref{QNS0} and \eqref{QNS1} can be rewritten as 
$$
\begin{aligned}
\bk{\Lambda[\cE] p_j\Theta[B_j] g} 
&= -\sum_{k=0}^{\infty} \sum_{l=0}^{\infty} \gamma_{kl} \sum_{|\alpha|=2k+1}\sum_{|\beta|=2l+1} \frac{1}{\alpha! \, \beta!} 
\bk{p_j(\nabla_{\pp}^\alpha \cE) \nabla_{\xx}^\alpha ((\nabla_{\xx}^\beta B_j)\nabla_{\pp}^\beta g)} 
\\
&= -\sum_{k=0}^{\infty} \sum_{l=0}^{\infty} \gamma_{kl} \sum_{|\alpha|=2k+1}\sum_{|\beta|=2l+1} \frac{1}{\alpha! \, \beta!} \,  \bk{p_j(\nabla_{\pp}^\alpha \cE) \nabla_{\pp}^\beta (\nabla_{\xx}^\alpha((\nabla_{\xx}^\beta B_j)g))} 
\\
&= -\sum_{k=0}^{\infty} \sum_{l=0}^{\infty} \gamma_{kl} \sum_{|\alpha|=2k+1}\sum_{|\beta|=2l+1} \frac{1}{\alpha! \, \beta!} \, \nabla_{\xx}^\alpha((\nabla_{\xx}^\beta B_j) \bk{p_j(\nabla_{\pp}^\alpha \cE) \nabla_{\pp}^\beta g)}),
\end{aligned}
$$
$$
\begin{aligned}
&\bk{\Lambda[\cE] \Theta_+[B_j] \frac{\partial g}{\partial x_j}} =
\\
&\qquad\qquad\quad = \sum_{k=0}^{\infty} \sum_{l=0}^{\infty} \gamma_{kl} \sum_{|\alpha|=2k+1}\sum_{|\beta|=2l} \frac{1}{\alpha! \, \beta!} 
\bk{(\nabla_{\pp}^\alpha \cE) \nabla_{\xx}^\alpha \left((\nabla_{\xx}^\beta B_j)\nabla_{\pp}^\beta \left(\frac{\partial g}{\partial x_j}\right)\right)} 
\\
&\qquad\qquad\quad = \sum_{k=0}^{\infty} \sum_{l=0}^{\infty} \gamma_{kl} \sum_{|\alpha|=2k+1}\sum_{|\beta|=2l} \frac{1}{\alpha! \, \beta!} 
\nabla_{\xx}^\alpha \left( (\nabla_{\xx}^\beta B_j) \frac{\partial}{\partial x_j}\bk{(\nabla_{\pp}^\alpha \cE) \nabla_{\pp}^\beta g}\right),
\end{aligned}
$$
$$
\begin{aligned}
\bk{\Lambda[\cE] \Theta[A^*] g} &= -\sum_{k=0}^{\infty} \sum_{l=0}^{\infty} \gamma_{kl} \sum_{|\alpha|=2k+1}\sum_{|\beta|=2l+1} \frac{1}{\alpha! \, \beta!} 
\bk{(\nabla_{\pp}^\alpha \cE) \nabla_{\xx}^\alpha \left((\nabla_{\xx}^\beta A^*)\nabla_{\pp}^\beta g\right)} 
\\
&=-\sum_{k=0}^{\infty} \sum_{l=0}^{\infty} \gamma_{kl} \sum_{|\alpha|=2k+1}\sum_{|\beta|=2l+1} \frac{1}{\alpha! \, \beta!}  \nabla_{\xx}^\alpha \left((\nabla_{\xx}^\beta A^*) \bk{(\nabla_{\pp}^\alpha \cE) \nabla_{\pp}^\beta g}\right),
\end{aligned}
$$
$$
\begin{aligned}
&\bk{\Lambda[\cE] p_i \, p_j\Theta[B_j] g} = 
\\
&\qquad\qquad = -\sum_{k=0}^{\infty} \sum_{l=0}^{\infty} \gamma_{kl} \sum_{|\alpha|=2k+1}\sum_{|\beta|=2l+1} \frac{1}{\alpha! \, \beta!} 
\bk{p_i \,p_j(\nabla_{\pp}^\alpha \cE) \nabla_{\xx}^\alpha ((\nabla_{\xx}^\beta B_j)\nabla_{\pp}^\beta g)}
\\
&\qquad \qquad = -\sum_{k=0}^{\infty} \sum_{l=0}^{\infty} \gamma_{kl} \sum_{|\alpha|=2k+1}\sum_{|\beta|=2l+1} \frac{1}{\alpha! \, \beta!} \, \nabla_{\xx}^\alpha ((\nabla_{\xx}^\beta B_j) \bk{p_i \,p_j(\nabla_{\pp}^\alpha \cE) \nabla_{\pp}^\beta g}),
\end{aligned}
$$
$$
\begin{aligned}
&\bk{\Lambda[\cE] p_i \Theta_+[B_j] \frac{\partial g}{\partial x_j}} = 
\\
&\qquad\qquad\quad =\sum_{k=0}^{\infty} \sum_{l=0}^{\infty} \gamma_{kl} \sum_{|\alpha|=2k+1}\sum_{|\beta|=2l} \frac{1}{\alpha! \, \beta!} 
\bk{p_i \,(\nabla_{\pp}^\alpha \cE) \nabla_{\xx}^\alpha \left((\nabla_{\xx}^\beta B_j)\nabla_{\pp}^\beta \left(\frac{\partial g}{\partial x_j}\right)\right)} 
\\
&\qquad\qquad\quad = \sum_{k=0}^{\infty} \sum_{l=0}^{\infty} \gamma_{kl} \sum_{|\alpha|=2k+1}\sum_{|\beta|=2l} \frac{1}{\alpha! \, \beta!} 
\nabla_{\xx}^\alpha \left( (\nabla_{\xx}^\beta B_j) \frac{\partial}{\partial x_j}\bk{(p_i \,\nabla_{\pp}^\alpha \cE) \nabla_{\pp}^\beta g}\right),
\end{aligned}
$$
$$
\begin{aligned}
\bk{\Lambda[\cE] p_i\Theta[A^*] g} &=-\sum_{k=0}^{\infty} \sum_{l=0}^{\infty} \gamma_{kl} \sum_{|\alpha|=2k+1}\sum_{|\beta|=2l+1} \frac{1}{\alpha! \, \beta!} 
\bk{p_i \, (\nabla_{\pp}^\alpha \cE) \nabla_{\xx}^\alpha \left((\nabla_{\xx}^\beta A^*)\nabla_{\pp}^\beta g\right)}  
\\
&=-\sum_{k=0}^{\infty} \sum_{l=0}^{\infty} \gamma_{kl} \sum_{|\alpha|=2k+1}\sum_{|\beta|=2l+1} \frac{1}{\alpha! \, \beta!}  \nabla_{\xx}^\alpha \left((\nabla_{\xx}^\beta A^*) \bk{p_i \, (\nabla_{\pp}^\alpha \cE) \nabla_{\pp}^\beta g}\right).
\end{aligned}
$$
Collecting these expressions gives Eqs.\ \eqref{NavSto1}) and \eqref{NavSto2}.

%\section*{Acknowledgements}

\bibliographystyle{siamplain}
\bibliography{bib_QHD}

\end{document}